\documentclass[%
 reprint,
 amsmath,amssymb,
 aps,
]{revtex4-2}

\usepackage{graphicx}
\usepackage{dcolumn}
\usepackage{subfigure}
\usepackage[normalem]{ulem}
\usepackage[dvipsnames]{xcolor}
\usepackage{bm}

\begin{document}

\preprint{APS/123-QED}

%
%
%
%
\date{\today}
%
%
%
\title{Numerical Simulations of a Superwalking Droplet}

\author{Rahil N. Valani$^{1}$}\email{rahil.valani@gmail.com}
\author{Tapio P. Simula$^{4}$}
\author{Anja C. Slim$^{2,3}$}%

\affiliation{$^1$School of Physics and Astronomy, Monash University, Victoria 3800, Australia}
\affiliation{$^2$School of Mathematics, Monash University, Victoria 3800, Australia}
\affiliation{$^3$School of Earth, Atmosphere and Environment, Monash University, Victoria 3800, Australia}
\affiliation{$^4$Optical Sciences Centre, Swinburne University of Technology, Melbourne 3122, Australia}

\begin{abstract}
A new class of walking droplets, coined superwalkers, has been observed when a bath of silicone oil is vibrated simultaneously at a given frequency and its subharmonic tone with a relative phase difference between them~\cite{superwalker}. In this paper, we present the details of the numerical implementation of a theoretical model for superwalkers that was developed by Valani \emph{et al.}~\cite{emersup}. The numerical analysis presented here provides the foundation for future numerical studies of superwalking droplets. 
\end{abstract}
\maketitle

\section{Introduction}
If a bath of silicone oil is vibrated vertically at frequency $f$, then a droplet of the same oil can be made to bounce indefinitely on the free surface of the liquid~\cite{Walker1978,Couder2005}. Increasing the driving amplitude results in the steady bouncing motion of the droplet becoming unstable and the droplet transitions to a walking state~\cite{Couder2005WalkingDroplets}. The walking droplet, also called a \emph{walker}, emerges when the driving acceleration is just below the Faraday instability threshold~\cite{Faraday1831a}, above which the whole surface becomes unstable to standing Faraday waves oscillating at the subharmonic frequency $f/2$. On each bounce, the walker generates a localised damped Faraday wave on the fluid surface. It then interacts with these waves on subsequent bounces, giving rise to a self-propelled wave-droplet entity. At high driving amplitudes below the Faraday threshold, the waves generated by the droplet decay very slowly. In this regime, the droplet is not only influenced by the wave it generated on its previous bounce, but also by the waves it generated in the distant past, giving rise to path memory in the system~\cite{Eddi2011}. In the high memory regime, walkers have been shown to mimic several peculiar behaviours that were previously though to be exclusive to the quantum regime. A detailed review of hydrodynamic quantum analogues of walking droplets is provided by Bush~\cite{Bush2015} and Bush \emph{et al.}~\cite{Bush2018}.

\begin{figure}
\centering
\includegraphics[width=\columnwidth]{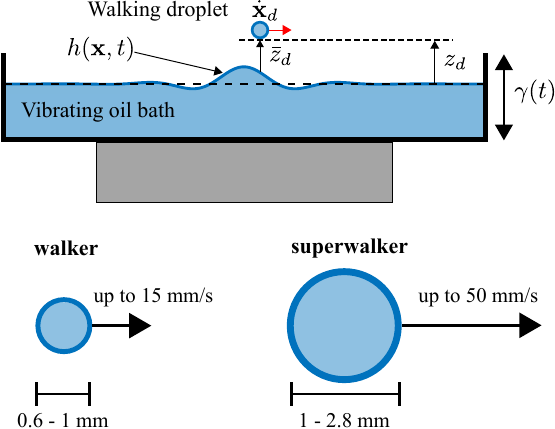}
\caption{(a) Schematic of the theoretical setup showing a droplet bouncing and walking on the surface of a vertically vibrated bath of the same liquid driven with acceleration $\gamma(t)=\gamma_f\sin(2\pi f t)+\gamma_{f/2}\sin(\pi f t + \Delta\phi)$. (b) Comparison of the size and speed of single-frequency ($f$) driven walkers with two-frequency ($f$ and $f/2$) driven superwalkers.}
\label{fig: schme}
\end{figure}

 Recently, a new class of walking droplets, coined \emph{superwalkers}, have been observed in experiments~\cite{superwalker}. These emerge when the bath is driven simultaneously at two frequencies, $f$ and $f/2$, with a relative phase difference $\Delta\phi$. For a commonly studied system with silicone oil of $20\,$cSt viscosity, driving the bath at a single frequency of $f=80\,$Hz produces walkers with diameters between $0.6\,$mm and $1\,$mm and walking speeds up to $15\,$mm/s~\cite{Molacek2013DropsTheory,Wind-Willassen2013ExoticDroplets}. In the same system with two frequency driving at $f=80$\,Hz and $f/2=40$\,Hz, superwalkers can be significantly larger than walkers with diameters up to $2.8\,$mm and walking speeds up to $50\,$mm/s~\cite{superwalker} (see bottom panel of figure~\ref{fig: schme}).  Moreover, the phase difference $\Delta\phi$ was observed to play a crucial role for superwalking droplets~\cite{superwalker} with peak superwalking speed occurring near $\Delta\phi=140^{\circ}$, while near $\Delta\phi=45^{\circ}$ the droplets only bounce or may even coalesce. 
 
 By using sophisticated numerical simulations that solve the detailed vertical and horizontal dynamics as well as the evolution of the free surface waves generated by the droplet, Galeano-Rios \emph{et al.}~\cite{galeano-rios_milewski_vanden-broeck_2019} were able to replicate superwalking behaviour for a single droplet of moderate radius $R=0.68$\,mm. They reported a good match in the superwalking speed between their simulation and the experiments of Valani \emph{et al.}~\cite{superwalker}. By doing a simple extension of an intermediate complexity model developed for walkers by Mol{\'a}{\v{c}}ek and Bush~\cite{molacek_bush_2013,Molacek2013DropsTheory} to two-frequency driven superwalkers, Valani \emph{et al.}~\cite{emersup} were able to capture the superwalking behaviour and reported good match between theory and experiments for small to medium sized superwalkers. In this paper, we provide details of the numerical implementation of the theoretical model for superwalkers developed by Valani \emph{et al.}~\cite{emersup} and we also present results of the effect of time step and memory of the droplet's wave field on the superwalking behaviour. 
 
\section{Theoretical formulation}

As shown schematically in figure~\ref{fig: schme}, consider a liquid droplet of mass $m$ and radius $R$ walking on a bath of the same liquid of density $\rho$, viscosity $\nu$ and surface tension $\sigma$. The bath is vibrating vertically with acceleration $\gamma(t)=\gamma_f\sin(2\pi f t)+\gamma_{f/2}\sin(\pi f t+\Delta\phi)$, where $\gamma_f$ is the amplitude of the primary driving frequency, $\gamma_{f/2}$ is the amplitude of the subharmonic frequency and $\Delta\phi$ is the relative phase difference between the two. The system geometry is described in the oscillating frame of the bath by horizontal coordinates $\mathbf{x} = (x,y)$ and the vertical coordinate $z$ with the origin chosen to be the undeformed surface of the bath (dashed horizontal line). In the comoving frame of the bath, the centre of mass of the droplet is located at a horizontal position $\mathbf{x}_d$ and the south pole of the droplet at a vertical position $z_d$ such that $z_d=0$ represents initiation of droplet's impact with the undeformed surface of the bath. The free surface elevation of the liquid filling the bath is at $z=h(\mathbf{x},t)$. Valani \emph{et al.}~\cite{emersup} developed a theoretical model for superwalkers by extending the model for walkers developed by Mol{\'a}{\v{c}}ek and Bush~\cite{molacek_bush_2013,Molacek2013DropsTheory} to two frequency driving. Here we review this model before proceeding to its numerical implementation.

\subsection{Vertical Dynamics}

To model the vertical motion, we use the linear-spring model of Mol{\'a}{\v{c}}ek and Bush~\cite{molacek_bush_2013} that results in the following equation of motion in the vertical direction
\begin{equation}\label{eq: vertical}
    m\ddot{z}_d=-m[g+\gamma(t)]+{F_N}(t).
\end{equation}
In this equation, the first term on the right hand side is the effective gravitational force on the droplet in the oscillating frame of the bath, with $g$ the constant acceleration due to gravity. The second term on the right hand side is the normal force imparted to the droplet during contact with the liquid surface. This contact force is calculated by modelling the bath as a spring and damper~\cite{molacek_bush_2013}, 
\begin{equation*}
{F_N}(t)=H(-\bar{z}_d) \,\max\left(-k\bar{z}_d-b\dot{\bar{z}}_d,0\right).
\label{springdamp}
\end{equation*}
Here, $H$ stands for the Heaviside step function and $\bar{z}_d=z_d-h(\mathbf{x}_d,t)$ is the height of the droplet above the free surface of the bath. The constants $k$ and $b$ are the spring constant and damping force coefficient, respectively. The corresponding dimensionless parameters are given by $K=k/m\omega_d^2$ and $B=b/m\omega_d$, where $\omega_d=\sqrt{\sigma/\rho R^3}$ is the droplet's characteristic oscillation frequency~\cite{molacek_bush_2013}.

\subsection{Wave field}

The free surface elevation $z=h(\mathbf{x},t)$ is calculated by adding individual waves generated by the droplet on each bounce:
\begin{equation}\label{eq: wave}
\textstyle h(\mathbf{x},t)=\sum_{n} h_n(\mathbf{x},\mathbf{x}_n,t,t_n)\, ,
\end{equation}
where $h_n(\mathbf{x},\mathbf{x}_n,t,t_n)$ is the wave field generated by bounce $n$ at location $\mathbf{x}_n$ and time $t_n$. 
The individual waves generated by the droplet on each bounce are localised decaying Faraday waves. For single frequency walkers driven at frequency $f$, these waves are subharmonic of wavelength $\lambda_F$~\cite{kumar_tuckerman_1994,Molacek2013DropsTheory,tadrist_shim_gilet_schlagheck_2018}. For two-frequency driving, Faraday waves have the same structure provided that $\gamma_f$ is dominant~\cite{PhysRevLett.71.3287}, and this is what has been also observed for superwalkers in experiments~\cite{superwalker}. Hence, we approximate the wave field generated by a superwalker using the wave field of a walker that has been modelled as a zeroth-order Bessel function that decays in time according to~\cite{Molacek2013DropsTheory,couchman_turton_bush_2019}
\begin{equation*}
h_n(\mathbf{x},\mathbf{x}_n,t,t_n)=AS\cos(\pi f t)\frac{\text{J}_0(k_F|\mathbf{x}-\mathbf{x}_n|)}{\sqrt{t-t_n}}\text{e}^{-(t-t_n)/T_F Me},
\end{equation*}
where $T_F=2/f$ is the Faraday period, $k_F=2\pi/\lambda_F$ is the Faraday wavenumber with $\lambda_F$ the Faraday wavelength and $Me=T_d/T_F(1-\gamma_f/\gamma_F)$ is the memory parameter with wave decay time $T_d=1/(\nu_e k_F^2)$, effective kinematic viscosity $\nu_e$ and $\gamma_F$ the Faraday threshold for single frequency driving at frequency $f$. The location and instant of the droplet's impact are approximated respectively by
\begin{equation*}
    \mathbf{x}_n=\int_{t_n^i}^{t_n^c}\mathbf{x}_d(t')F_N(t')\,\text{d}t'\Big{/}\int_{t_n^i}^{t_n^c}F_N(t')\,\text{d}t',
\end{equation*}
and
\begin{equation*}
    t_n=\int_{t_n^i}^{t_n^c}t'F_N(t')\,\text{d}t'\Big{/}\int_{t_n^i}^{t_n^c}F_N(t')\,\text{d}t',
\end{equation*}
where $t_n^i$ and $t_n^c$ are the time of initiation and completion of the $n$th impact. The wave-amplitude coefficient $A$ and impact-phase parameter $S$ are given by
\begin{equation*}
    A=\sqrt{\frac{2\nu_e}{\pi}}\frac{1}{\sigma}\frac{k_F^3 R^2}{3 k_F^ 2R^2 + Bo}
\end{equation*}
and
\begin{equation*}
    S=\int_{t_n^i}^{t_n^c} F_N(t')\sin(\pi f t') dt'.
\end{equation*}

where $Bo=\rho g R^2/\sigma$ is the Bond number. 

\subsection{Horizontal Dynamics}

To model the horizontal dynamics, we use the model of Mol{\'a}{\v{c}}ek and Bush~\cite{Molacek2013DropsTheory} that results in the following equation of motion in the horizontal direction
\begin{equation}\label{eq: hor}
    m\ddot{\mathbf{x}}_d+D_{tot}(t)\dot{\mathbf{x}}_d=-{F_N}(t) \boldsymbol{\nabla} h(\mathbf{x}_d,t),
\end{equation}
where $D_{tot}(t)=C\sqrt{\frac{\rho R}{\sigma}}{F_N}(t)+6\pi R \mu_a$ is the total instantaneous drag force coefficient, comprising of momentum loss during contact with the bath and air drag, respectively. Here $C$ is the contact drag coefficient and $\mu_a$ is the dynamic viscosity of air. The force on the right hand side is the horizontal component of the contact force arising from the small slope of the underlying wave field.
 
\section{Numerical implementation}

 \begin{figure*}
\centering
\includegraphics[width=2.12\columnwidth]{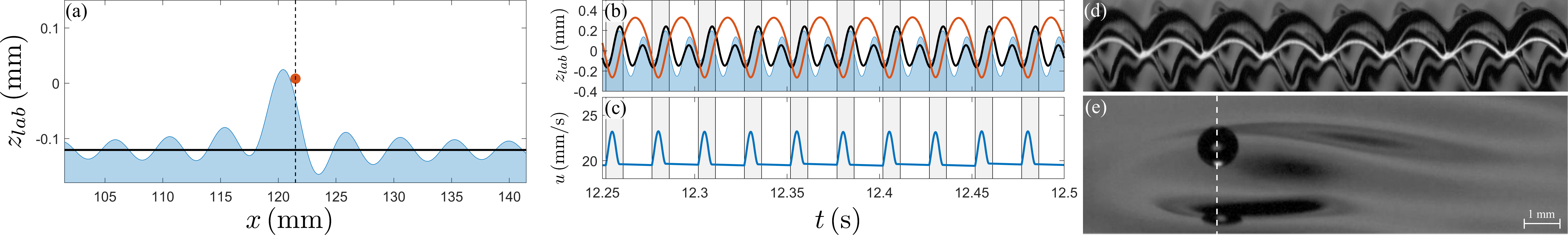}
\caption{Numerical simulation of a superwalker. (a) Instantaneous snapshot of a superwalker (the red circle represents the south pole of the droplet) just before it impacts the underlying wave field (filled blue region). The solid black horizontal line represents the vertical location of the undeformed surface of the bath. By taking a vertical slice of panel (a) along the centre of the droplet (dashed vertical line) and juxtaposing all such slices at different times, we get an evolution plot as shown in panel (b) of the vertical position of the bath, $\mathcal{B}(t)=-(\gamma_f/(2\pi f )^2) \sin(2\pi f t) - (\gamma_{f/2}/(\pi f)^2) \sin(\pi f t +\Delta \phi)$, the vertical height of the wave field beneath the droplet, $h(x_d,t)+\mathcal{B}(t)$, and the vertical motion of south pole of the droplet, $z_d(t)+\mathcal{B}(t)$, all in the lab frame. The filled grey region indicates contact between the droplet and the wave. Panel (c) shows the evolution of the horizontal walking velocity $u$. For this simulation, the timestep and the past number of waves were fixed to $\Delta t/T_F=1/250$ and $N=100$ respectively. Panels (d) and (e) show the vertical dynamics of the droplet (duration of $12$\,milliseconds) and a snapshot of the droplet from experiments of Valani \emph{et al.}~\cite{superwalker}. The plot in panel (d) is generated in a similar way to panel (b) by juxtaposing vertical slices (white dotted line in (e)) at different times from the experimental images. Here the parameter values for both the numerical simulation and experiments are fixed to $R=0.54\,$mm, $\gamma_{80}=3.8\,g$, $\gamma_{40}=0.6\,g$ and $\Delta\phi=130^{\circ}$.}
\label{fig: SWim}
\end{figure*}

Solitary walkers and superwalkers follow straight line trajectories~\cite{Oza2013,superwalker}. Hence to simulate a superwalker, we proceed by restricting the domain of horizontal motion to $x$ direction only. To solve this system numerically, we discretise equations~(\ref{eq: vertical}) and (\ref{eq: hor}) using the Leap-frog method~\cite{Sprott}, a modified version of the Euler method where the new horizontal and vertical positions are calculated using the old velocities and then the new velocities are calculated using the new positions. Converting the second order differential equation for the vertical dynamics in equation~(\ref{eq: vertical}) into a system of two first order ordinary differential equations and discretising using the Leap-frog method we get,

\begin{equation*}\label{eq: Vertical disc 1}
    z_d(t_{i+1})=z_d(t_{i})+\Delta t v_d(t_{i}),
\end{equation*}
and
\begin{equation*}\label{eq: Vertical disc 2}
    v_d(t_{i+1})=v_d(t_{i})+\frac{\Delta t}{m} \left[-m(g+\gamma(t_{i+1}))+F_N(t_{i+1})\right],
\end{equation*}
where $v_d(t)=\dot{z}_d(t)$ and
\begin{equation*}\label{eq: Vertical disc 3}
{F_N}(t_{i+1})=H(-\bar{z}_d(t_{i+1})) \,\max\left(-k\bar{z}_d(t_{i+1})-b\bar{v}_d(t_{i}),0\right).
\end{equation*}
Here $\bar{z}_d(t_{i+1})=z_d(t_{i+1})-h(x_d(t_{i+1}),t_{i+1})$ and $\bar{v}_d(t_{i})=v_d(t_{i})-\frac{\partial h}{\partial t}(x_d(t_{i+1}),t_{i+1})$. The total wave height beneath the droplet $h(x_d(t_{i+1}),t_{i+1})$ is calculated using equation~(\ref{eq: wave}) by keeping the waves from last $N$ impacts of the droplet. The integral required to calculate the location of impact $\mathbf{x}_n$, the time of impact $t_n$ and the impact-phase parameter $S$ were done using the MATLAB inbuilt trapezoid function. Similarly, the second order equation~(\ref{eq: hor}) governing the horizontal dynamics takes the following form,
\begin{equation*}\label{eq: Horizontal disc 1}
    x_d(t_{i+1})=x_d(t_{i})+\Delta t u_d(t_{i}),
\end{equation*}
and
\begin{align*}\label{eq: Horizontal disc 2}
    u_d(t_{i+1})=u_d(t_{i})+\frac{\Delta t}{m}\Big[-D_{tot}(t_{i+1})u_d(t_{i})&\nonumber\\
    -{F_N}(t_{i+1}) \frac{\partial h}{\partial x}(x_d(t_{i+1}),t_{i+1})\Big]&,
\end{align*}
where $u_d(t)=\dot{x}_d(t)$.

 We fix the physical parameters to match the experiments of Valani \emph{et al.}~\cite{superwalker}: $\rho=950$\,kg/m$^3$, $\nu=20$\,cSt, $\sigma=20.6$\,mN/m, $\gamma_F=4.2\,g$, $\lambda_F=5$\,mm and $f=80$\,Hz. There are three adjustable parameters whose values are not known for superwalkers: the dimensionless spring constant of the bath $K$, the dimensionless damping coefficient of the bath $B$ and the contact drag coefficient $C$. For walkers, the typical values used for these parameters are $K=0.59$ and $B=0.48$ \cite{couchman_turton_bush_2019}, and $C=0.17$ \cite{Molacek2013DropsTheory}. For superwalkers, we also take $C=0.17$ but adjust $K$ and $B$ to $0.8$ as it gives the best match with experimental data. Unless otherwise stated, we also fix the following parameters: $R=0.54\,$mm, $\gamma_{80}=3.8\,g$, $\gamma_{40}=0.6\,g$, $\Delta\phi=130^{\circ}$, $\Delta t = T_F/250$ and $N=100$ past impacts. The simulations are initialised with $x_d=0$\,mm, $u_d=1\,$mm/s, $v_d=0$\,mm/s and $z_d=0$\,mm.

\section{Results and Discussion}

 Figure~\ref{fig: SWim}(a-c) shown the evolution of the vertical dynamics and the horizontal walking speed of a simulated superwalker. A snapshot of the simulation is shown in figure~\ref{fig: SWim}(a) while an evolution plot of the vertical motion of the droplet, its underlying wave and the bath is shown in figure~\ref{fig: SWim}(b). Here it can be seen that the droplet skips every second peak in the bath motion, in agreement with the bouncing motion observed experimentally by Valani \emph{et al.}~\cite{superwalker} at the same parameter values (see figure~\ref{fig: SWim}(d) and (e)). Figure~\ref{fig: SWim}(c) shows the horizontal velocity of the droplet $u$ as a function of time $t$. When the droplet is not in contact with the bath, its horizontal speed decreases slowly due to the air drag experience by the droplet. While in contact with the bath, we see that the droplet accelerates and then decelerates horizontally, dictated by the horizontal component of the contact force acting on the droplet. From this time evolution we obtain an average walking speed of $\bar{u}\approx20\,$mm/s which agrees well with the experimentally observed average speed of approximately $21\,$mm/s at these parameter values~\cite{superwalker}.

 \begin{figure}
\centering
\includegraphics[width=\columnwidth]{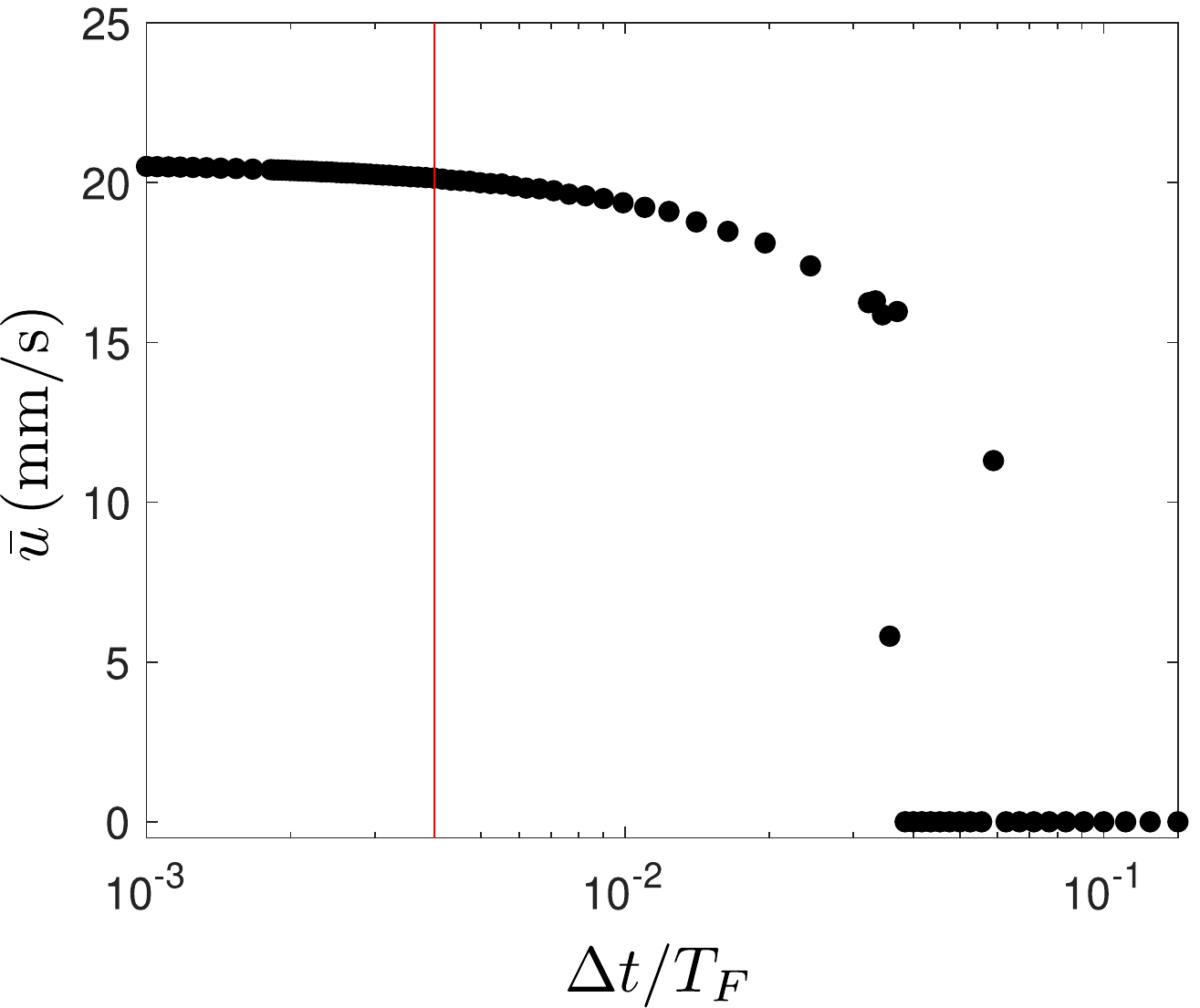}
\caption{Average walking speed $\bar{u}$ as a function of the scaled time step $\Delta t/T_F$ for parameter values $R=0.54\,$mm, $\gamma_{80}=3.8\,g$, $\gamma_{40}=0.6\,g$, $\Delta\phi=130^{\circ}$ and $N=100$. The red vertical line indicates the time step $\Delta t/T_F=1/250$.}
\label{fig: speedtimestep}
\end{figure}

We investigate the convergence of the simulations by simulating the superwalker using different time steps $\Delta t$ and plotting the average walking speed $\bar{u}$ as a function of $\Delta t$ scaled by the Faraday period $T_F$ (see figure~\ref{fig: speedtimestep}). In the periodic vertical motion depicted in figure~\ref{fig: SWim}(b), the period $T_F$ also corresponds to the period of the bouncing motion. We find that for $\Delta t/T_F \gtrsim 0.14$, the simulation do not converge. For $0.03 \lesssim \Delta t/T_F \lesssim 0.14$, we do not capture the correct vertical dynamics and hence we do not see any horizontal walking motion. For $\Delta t/T_F \lesssim 0.03$, we capture the correct walking behaviour. Our chosen time step of $\Delta t/T_F=1/250$ (red line) used for the results presented in figure~\ref{fig: SWim} has converged with an average droplet speed of $\bar{u}\approx20$\,mm/s. 

\begin{figure}
\centering
\includegraphics[width=\columnwidth]{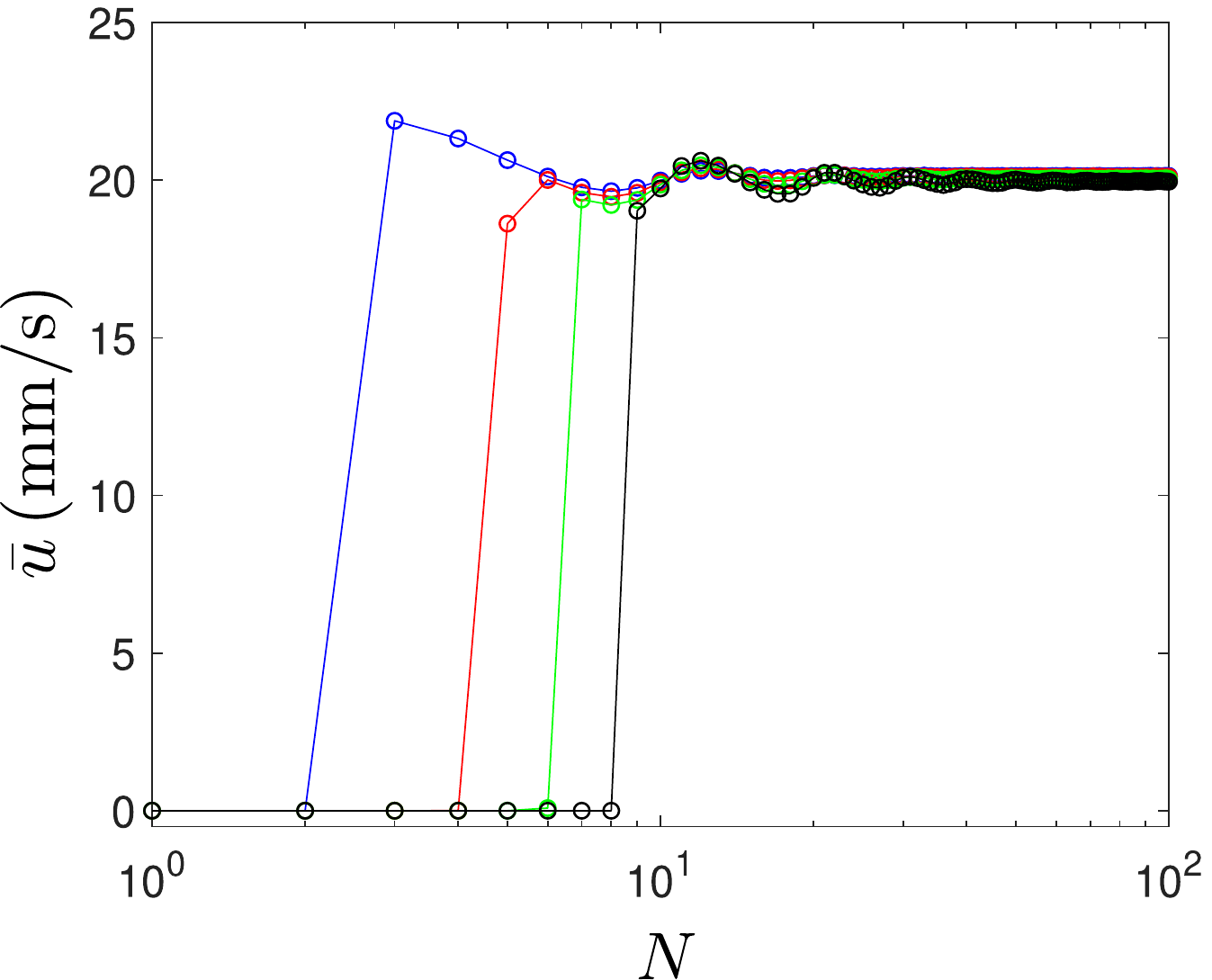}
\caption{Average walking speed $\bar{u}$ as a function of the past number of waves $N$ incorporated in the total wave field for four different parameter values $\gamma_{80}=3.8\,g$~(blue),~$3.9\,g$~(red),~$4.0\,g$~(green) and $4.1\,g$~(black). Other parameter values are fixed to $R=0.54\,$mm, $\gamma_{40}=0.6\,g$, $\Delta\phi=130^{\circ}$ and $\Delta t/T_F=1/250$.}
\label{fig: past waves}
\end{figure}

To understand the role of memory in the simulation of a superwalking droplet, we simulated the superwalker by varying the number of past impacts $N$ that are incorporated in the droplet's total wave field for difference values of the acceleration amplitude $\gamma_{80}$. A plot of the average walking speed $\bar{u}$ as a function of the number of past impacts taken $N$ for four different $\gamma_{80}$ values is shown in figure~\ref{fig: past waves}. As expected, the number of past waves required to capture the superwalking behaviour increases as $\gamma_{80}$ increases, since the waves at higher $\gamma_{80}$ decay more slowly. However, we find that incorporating as little as $3$ past waves for $\gamma_{80}=3.8\,g$ and $9$ past waves for $\gamma_{80}=4.1\,g$ are enough to capture the correct bouncing mode and get the average walking speed $\bar{u}$ within $10\%$ of its converged value.

Single-frequency driven walkers have been shown to mimic several quantum analogues in the high memory regime in both experiments and numerical simulations~\cite{Bush2015}. Thus, it would be interesting to revisit these experiments using superwalkers. Superwalkers provide an extra degree of freedom where the phase difference between the two driving signals can be used to tune the speed of the superwalkers~\cite{superwalker}. Such investigations may give us new insights into the role of inertia in hydrodynamic quantum analogues. The numerical analysis presented here provides the foundation to perform such numerical experiments with superwalkers.

\section{Conclusion}
In this paper, we provided details of the numerical method to simulate superwalking droplets using the theoretical model presented in Valani \emph{et al.}~\cite{emersup}. We have shown that the bouncing modes and the average walking speed agrees well with experiments. By varying the number of past waves incorporated in the wave field, we found that only a few past waves are enough to capture the correct bouncing mode with the average walking speed being within $10\%$ of the converged solution. The numerical method provided here lays a strong foundation for future numerical studies of superwalking droplets.

\section{Acknowledgements}
We acknowledge financial support from an Australian Government Research Training Program (RTP) Scholarship (R.V.) and the Australian Research Council via the Future Fellowship Project No. FT180100020 (T.S.).

\bibliography{Superwalker_numerical}

\end{document}